\input{aipcheck}
\documentclass[final]{aipproc}
\layoutstyle{6x9}

\begin{document}

\title{Constraints on SUSY Seesaw from Leptogenesis and LFV}



\author{Frank Deppisch}{address={Deutsches Elektronen-Synchrotron DESY,
D-22603 Hamburg, Germany}}
\author{Simon Albino}{address={Deutsches Elektronen-Synchrotron DESY,
D-22603 Hamburg, Germany}}

\author{Reinhold R\"uckl}{address={Institut f\"ur Theor.
Physik und Astrophysik,
Universit\"at W\"urzburg, D-97074 W\"urzburg, Germany}}

\begin{abstract}
We study constraints on the fundamental parameters of
supersymmetric type I seesaw models imposed by neutrino data,
charged lepton flavor violation, thermal leptogenesis and
perturbativity of the neutrino Yukawa couplings.
\end{abstract}

\maketitle

\section{SUSY Seesaw Model and slepton mass matrix}
The seesaw mechanism involving heavy right-handed Majorana
neutrinos can naturally explain the smallness of the neutrino
masses. A consequence in supersymmetric extensions of the standard
model, which may be tested by experiment, is the violation of
lepton flavor conservation (LFV). In addition, heavy Majorana
neutrinos violate lepton number which via leptogenesis may give
rise to the observed baryon-antibaryon asymmetry of the universe.

If three right-handed neutrino singlet fields $\nu_R$ are added to
the MSSM particle content, one gets additional terms in the
superpotential,
\begin{equation}
\label{suppot4}
    W_\nu = -\frac{1}{2}\nu_R^{cT} M \nu_R^c + \nu_R^{cT} Y_\nu L
    \cdot H_2,
\end{equation}
where $L$ and $H_2$ denote the left-handed lepton and hypercharge
+1/2 Higgs doublets, respectively, \(Y_\nu\) is the matrix of
neutrino Yukawa couplings and $M$ is the neutrino Majorana mass
matrix. If the Majorana mass scale is much greater than the
electroweak scale, and thus much greater than the scale of the
Dirac mass matrix \(m_D=Y_\nu \langle H_2^0 \rangle\), \(\langle
H_2^0 \rangle = v \sin\beta\) being the appropriate Higgs v.e.v.
with \(v=174\)~GeV and \(\tan\beta =\langle H_2^0\rangle/\langle
H_1^0\rangle\)), the resulting mass matrix $M_\nu$ for the light
neutrinos is given by
\begin{equation}
\label{eqn:SeeSawFormula}
    M_\nu = m_D^T M^{-1} m_D = Y_\nu^T M^{-1} Y_\nu (v \sin\beta )^2.
\end{equation}
This matrix is diagonalized by a unitary transformation,
\begin{equation}
\label{eqn:NeutrinoDiag}
    U_{MNS}^T M_\nu U_{MNS} = \textrm{diag}(m_1,m_2,m_3),
\end{equation}
leading to the physical mass eigenvalues $m_i$. The remaining
neutrino mass eigenstates are too heavy to be observed directly.
However through virtual corrections they induce small off-diagonal
terms in the renormalized MSSM slepton mass matrix,
\begin{eqnarray}
 m_{\tilde l}^2=\left(
    \begin{array}{cc}
        m_L^2    & (m_{LR}^{2})^\dagger \\
        m_{LR}^2 & m_R^2
    \end{array}
      \right)_{\rm MSSM}+\left(
    \begin{array}{cc}
       \delta m_L^2    & (\delta m_{LR}^{2})^\dagger \\
        \delta m_{LR}^2 & 0
    \end{array}
      \right),
\end{eqnarray}
which may give rise to observable LFV processes. In the minimal
supergravity scheme (mSUGRA) and in leading logarithmic
approximation, these corrections read \cite{Hisano:1999fj}
\begin{eqnarray}
\label{left_handed_SSB2}
    \delta m_{L}^2 = -\frac{1}{8 \pi^2}(3m_0^2+A_0^2)(Y_\nu^\dag L Y_\nu)
    ,\quad \quad
    \delta m_{LR}^2
       = -\frac{3 A_0 v \cos\beta}{16\pi^2}(Y_l Y_\nu^\dag L Y_\nu),
\end{eqnarray}
where $L_{ij} = \ln(M_{GUT}/M_i)\delta_{ij}$, $M_i$ being the
heavy neutrino mass eigenvalues, and \(m_0\) and \(A_0\) are the
common scalar mass and trilinear coupling, respectively, at the
unification scale $M_{GUT}$. The product $Y_\nu^\dagger L Y_\nu$
entering these corrections can be determined by making use of the
result~\cite{Casas:2001sr}
\begin{eqnarray}
\label{eqn:yy}
    Y_\nu \!=\!
        \frac{1}{v\sin\beta}\textrm{diag}(\sqrt{M_1},\sqrt{M_2},\sqrt{M_3})
        \!\cdot\!R\!\cdot\!\textrm{diag}(\sqrt{m_1},\sqrt{m_2},\sqrt{m_3})
        \!\cdot\! U_{MNS}^\dagger,
\end{eqnarray}
following from (\ref{eqn:SeeSawFormula}) and
(\ref{eqn:NeutrinoDiag}). The unknown complex orthogonal matrix
$R$ appearing in (\ref{eqn:yy}) can be parametrized in terms of 3
complex angles $\theta_i=x_i+i y_i$. Using neutrino data as input,
$Y_{\nu}$, respectively $Y_\nu^\dagger L Y_\nu$, is to be evolved
from the electroweak to the GUT scale.

\section{Charged lepton flavor violation}
\begin{figure}[t]
\centering
\includegraphics[clip,width=0.56\textwidth]{figs/low_SPS1.eps}
\includegraphics[clip,width=0.42\textwidth]{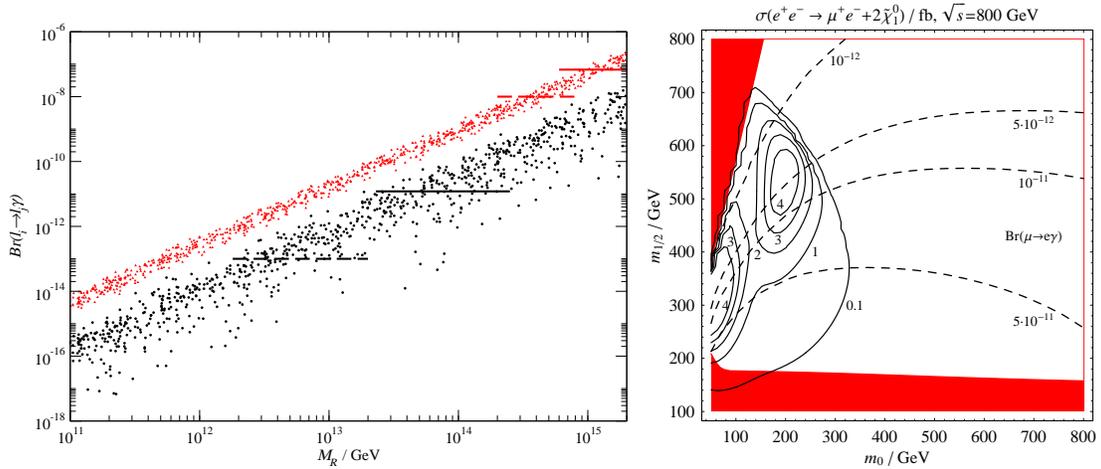}
\caption{[left] Br$(\tau\to\mu\gamma)$ (upper) and Br$(\mu\to
e\gamma)$ (lower) versus $M_R$ in mSUGRA scenario SPS1a
~\cite{Battaglia:2001zp} for real $R$. The solid (dashed)
horizontal lines mark existing bounds (expected sensitivities of
future experiments)~\cite{Deppisch:2003wt,PDG}.
[right] Contours of the polarized cross section
\(\sigma(e^+e^-\to\mu^+e^- +2\tilde\chi_1^0)\) (solid) and of
Br$(\mu\to e\gamma)$ (dashed) in the \(m_0-m_{1/2}\) plane. The
collider energy is \(\sqrt{s}_{ee}=800\)~GeV and the beam
polarizations are $P_{e^-} = +0.9, P_{e^+} = +0.7$. The remaining
mSUGRA parameters are chosen to be \(A_0=0\),\(\tan\beta=5\) and
\(\textrm{sign}(\mu)=+\). The neutrino oscillation parameters are
fixed at their best fit values~\cite{Gonzalez-Garcia:2001sq}, the
lightest neutrino mass \(m_1\) and all complex phases are set to
zero, and the degenerate heavy neutrino mass is taken to be
\(M_R=10^{14}\)~GeV. The shaded (red) regions are forbidden by
mass bounds from various experimental sparticle searches.}
\label{br12vsMRdegRSPS1a}
\end{figure}
For a first brief discussion, we assume the heavy Majorana
neutrinos to be degenerate in mass, $M_i=M_R$, and the matrix $R$
to be real, $y_i=0$. In this case, $R$ drops out from the product
$Y_\nu^\dagger L Y_\nu$ in (\ref{left_handed_SSB2}).

In the presence of the flavor off-diagonal terms
(\ref{left_handed_SSB2}) in the slepton mass matrix, virtual
slepton exchange in loops induces LFV radiative decays \(l_i\to
l_j \gamma\). To lowest order in \((\delta m_L)^2_{ij}\) one has
\cite{Hisano:1999fj,Casas:2001sr}
\begin{equation}
\label{eqn:DecayApproximation}
    \Gamma(l_i \to l_j \gamma) \propto \alpha^3 m_{l_i}^5
    \frac{|(\delta m_L^2)_{ij}|^2}{\tilde{m}^8} \tan^2 \beta,
\end{equation}
where $\tilde m$ characterizes the typical sparticle masses in the
loop.
Another feasible test of LFV is provided by the linear collider
processes $e^+e^-\to \tilde{l}_a^-\tilde{l}^+_b\to l_i^-l^+_j +
2\tilde{\chi}^0_1$. Analogously to (\ref{eqn:DecayApproximation}),
one finds the approximate cross section~\cite{Deppisch:2003wt}
\begin{equation}\label{eqn:HighEnergyApproximation}
    \sigma(e^+e^- \to l_i^- l_j^+ +2\tilde\chi^0_1)
    \approx
        \frac{|(\delta m_L^2)_{ij}|^2}{m^2_{\tilde l}
        \Gamma^2_{\tilde l}}
        \sigma(e^+e^- \rightarrow l_i^- l_i^+
        +2\tilde\chi^0_1)
\end{equation}
in the limit of small slepton mass corrections.

The left plot of Fig.~\ref{br12vsMRdegRSPS1a} displays Br$(\mu \to
e \gamma)$ and Br$(\tau \to \mu \gamma)$ as a function of the
Majorana mass \(M_R\)~\cite{Deppisch:2002vz} in the mSUGRA
scenario SPS1a~\cite{Battaglia:2001zp}. Also indicated are the
current bounds~\cite{PDG} and expected sensitivities of future
experiments~\cite{Deppisch:2003wt}. As can be seen, the existing
limit Br$(\mu\to e\gamma)<1.2 \times 10^{-11}$ implies the upper
bound \(M_R < 3\cdot10^{14}\)~GeV, while Br$(\mu\to
e\gamma)<10^{-13}$ would probe Majorana masses smaller by roughly
one order of magnitude. In comparison, the experimental
sensitivity in the channel $\tau \to \mu \gamma$ cannot quite
compete with $\mu \to e \gamma$, at least in the scenario
considered. On the other hand, Br$(\tau\to\mu\gamma)$ is less
affected by experimental errors in the light neutrino parameters,
allowing in principle a more accurate determination of $M_R$ than
Br$(\mu\to e\gamma)$.

The right plot of Fig.~\ref{br12vsMRdegRSPS1a} shows contours of
fixed values of the polarized cross section for
\(\sigma(e^+e^-\to\mu^+e^- +2\tilde\chi_1^0)\) and of Br$(\mu\to
e\gamma)$ in the \(m_0-m_{1/2}\) parameter plane. As expected, for
large $m_0$, resulting in relatively heavy sleptons, rare decay
experiments are superior probes of LFV, while for sufficiently
small $m_0$, linear collider experiments could probe regions of
large $m_{1/2}$ not yet excluded by the present bound on
Br$(\mu\to e\gamma)$.

\section{Leptogenesis}
Thermal leptogenesis is an efficient mechanism for generating the
observed baryon asymmetry of the universe (see e.g.
\cite{Buchmuller:2004nz}). As well known, simple realizations of
leptogenesis favor a hierarchical spectrum of heavy Majorana
neutrinos, $M_1 \ll M_2 \ll M_3$. Out-of-equilibrium decays of the
lightest species of the heavy neutrinos result in a lepton
asymmetry which is determined by the mass $M_1$, the light
neutrino masses and the imaginary parts of the matrix \(R\). Later
in the evolution of the universe the lepton asymmetry is converted
in a baryon asymmetry by sphaleron processes.
\begin{figure}[t]
\centering
\includegraphics[clip,width=0.49\textwidth]{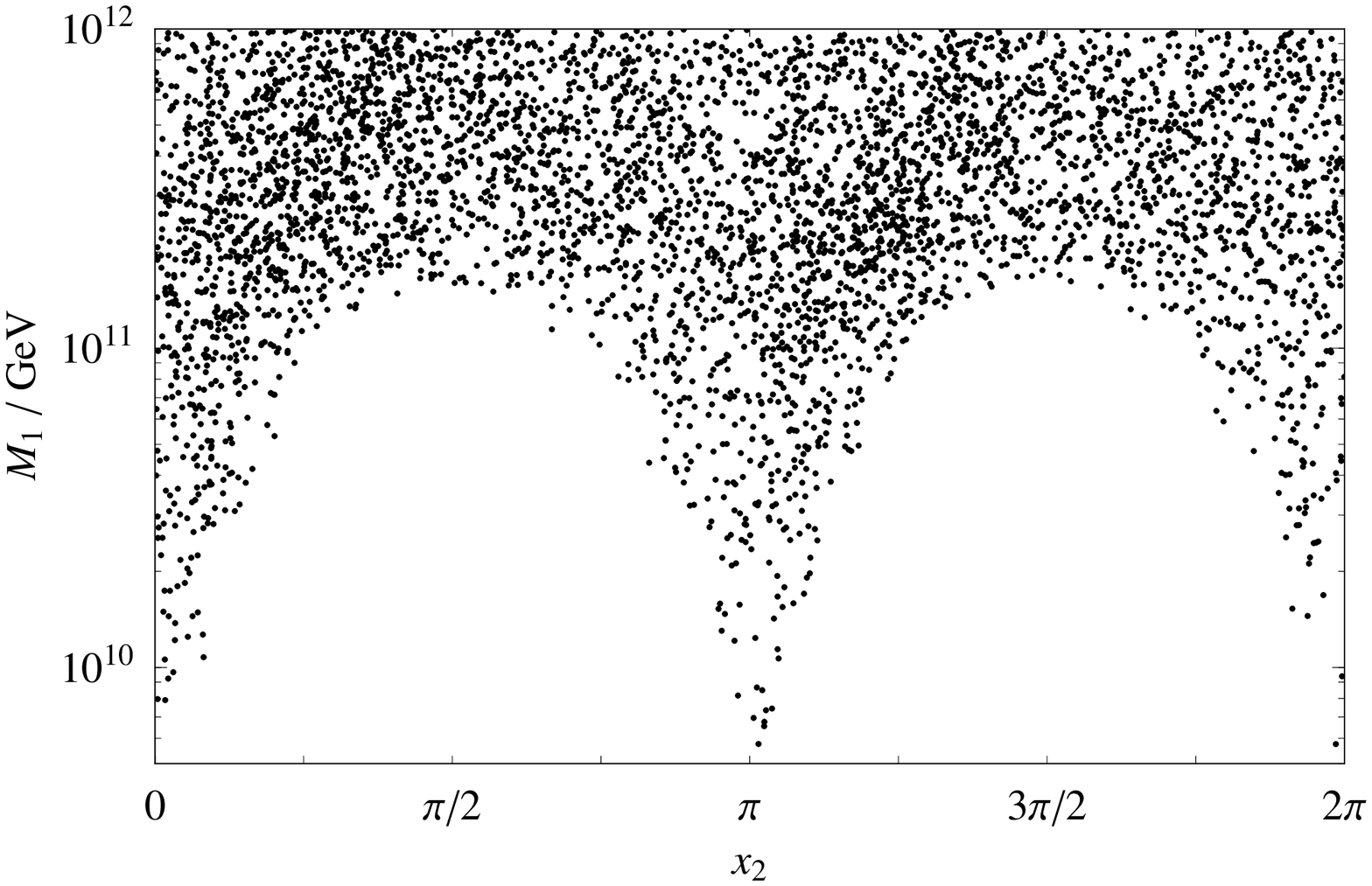}
\includegraphics[clip,width=0.49\textwidth]{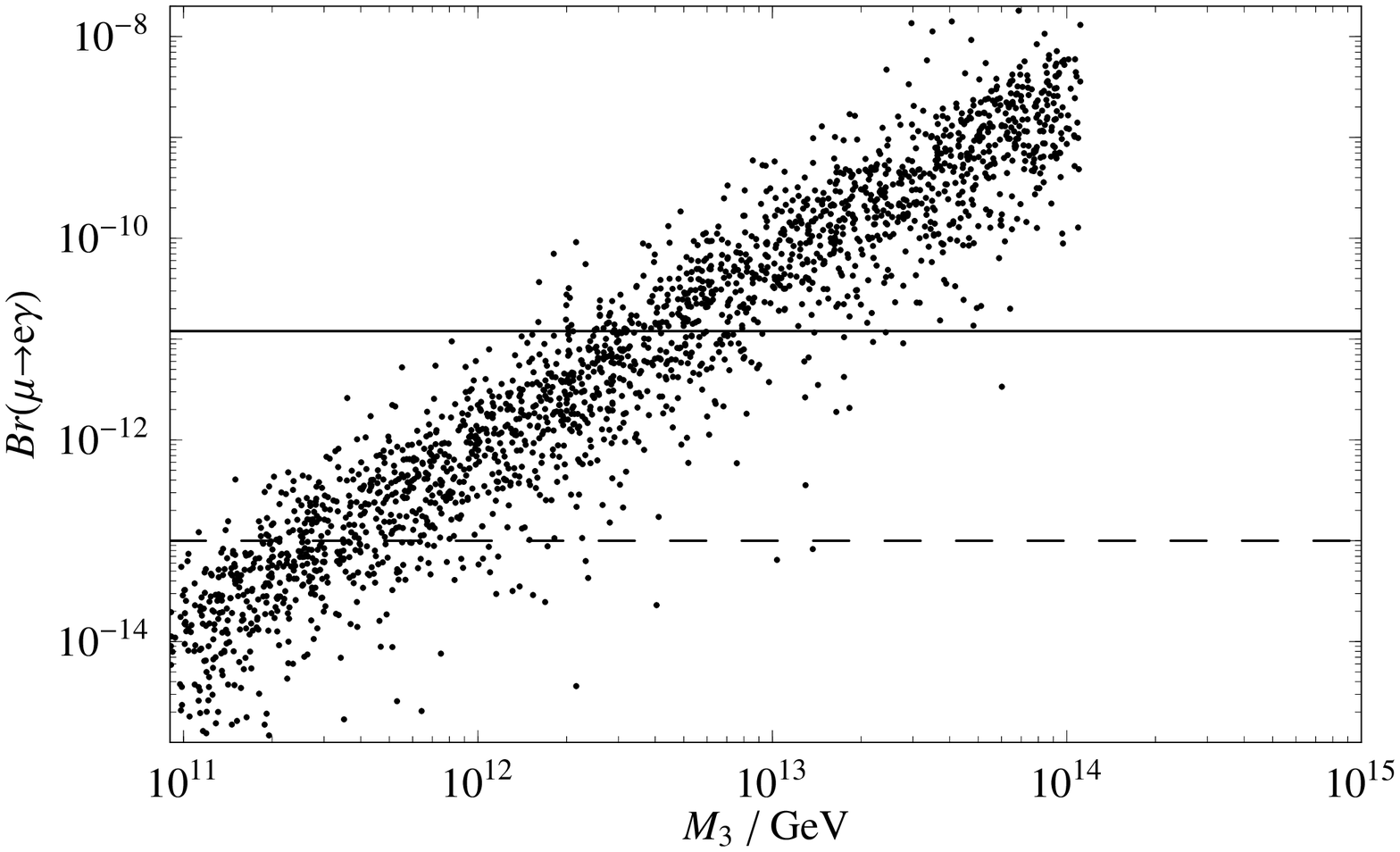}
\caption{[left] Values of $(M_1,x_2)$ consistent with the baryon
to photon ratio \(\eta_B=6.3\cdot 10^{-10}\) via leptogenesis.
[right] \(Br(\mu\to e\gamma)\) vs. $M_3$  for \(M_1=10^{10}\)~GeV
($|\cos \theta_2| \approx 1$) in the mSUGRA scenario
SPS1a~\cite{Battaglia:2001zp}. The solid (dashed) line indicates
the existing bound (expected MEG sensitivity). All other seesaw
parameters are scattered in their allowed ranges assuming
hierarchical light and heavy neutrinos.} \label{fig:torboegen}
\end{figure}
The condition to reproduce the observed baryon to photon ratio
\(\eta_B = 6.3\cdot 10^{-10}\) puts constraints on the parameters
\(x_2\) and \(x_3\) of the \(R\) matrix~\cite{Deppisch:2005rv}.
This is illustrated in the left plot of Fig.~\ref{fig:torboegen}
for $x_2$. One sees that for $M_1 < 10^{11}$~GeV, $x_2$ has to
approach the values $n\pi$. A similar behavior is found for $x_3$.

As discussed in the previous section for degenerate heavy
neutrinos and real $R$, measurements of Br$(l_i\to l_j\gamma)$ can
be used to constrain the heavy neutrino masses. This is
illustrated in the right plot of Fig.~\ref{fig:torboegen} for
$\mu.\to e \gamma$ and $M_3$. As can be seen, the present bound
Br$(\mu\to e \gamma) < 1.2 \times 10^{-11}$ constrains $M_3$ to be
smaller than about $10^{13}$~GeV. The sensitivity of the MEG
experiment at PSI will set the limit $M_3<{\cal O}(10^{12})$~GeV
or provide a first rough determination of $M_3$.

\begin{theacknowledgments}

We thank H. P\"as and Y. Shimizu for collaboration.
This work was supported by the German Federal Ministry of Education and Research
(BMBF) under contract 05HT6WWA.

\end{theacknowledgments}

\end{document}